\documentclass[prc,superscriptaddress,unsortedaddress,twocolumn]{revtex4}
\usepackage[english]{babel}
\usepackage{graphicx,epsf,bm,amsmath,color}
\usepackage{here,ulem}

\def\bk{\mbox{\boldmath $k$}}

\def\bq{\mbox{\boldmath $q$}}

\def\bfsigma{\mbox{\boldmath $\sigma$}}

\def\bftau{\mbox{\boldmath $\tau$}}
\def\h3t{\mbox{$_\Lambda^3$H}}

\begin{document}
\title{Faddeev calculation of \h3t incorporating 2$\pi$-exchange $\Lambda$NN interaction}
\author{H. Kamada}\email{kamada@mns.kyutech.ac.jp}
\affiliation{Department of Physics, Faculty of Engineering, Kyushu Institute of Technology,
Kitakyushu 804-8550, Japan}

\author{M. Kohno}\email{kohno@rcnp.osaka-u.ac.jp}
\affiliation{Research Center for Nuclear Physics, Osaka University, Ibaraki 567-0047,
Japan}

\author{K. Miyagawa}\email{miyagawa@rcnp.osaka-u.ac.jp}
\affiliation{Research Center for Nuclear Physics, Osaka University, Ibaraki 567-0047,
Japan}

\begin{abstract}
Faddeev calculations of hypertriton (\h3t) separation energy are performed,
incorporating $2\pi$-exchange $\Lambda $NN three-baryon force. Repulsive contributions of
the three-baryon force in the order of 20 keV are found, depending on the NN interactions
employed. The effect is not negligible compared with the small separation-energy of \h3t.
\end{abstract}

\maketitle
\section{Introduction}
The appropriate description of the interaction of the $\Lambda$ hyperon with nucleons is
essential to microscopically understand the existing mode of $\Lambda$ in nuclei
and dense neutron-star matter. Because experimental data of the
$\Lambda$-nucleon ($\Lambda$N) scattering is scarce at present, the description of the $\Lambda$N
potential has ambiguities. There is no two-body $\Lambda$-nucleon bound state, and
therefore the hypertriton holds a key position as the lightest
bound system to constrain the $\Lambda$N interactions. However, there is
a complexity associated with the appearance of three-baryon forces (3BFs).

When two-body interactions that are constructed by eliminating various degrees of
freedom are applied in many-body systems, three-body interactions should appear. 
The recent development of the construction of baryon-baryon interactions in chiral
effective field theory (ChEFT) \cite{EHM09,ME11} provides a systematic way
to introduce three-body (and more-than-three-body) forces in a power-counting
scheme and, therefore, quantifies the role of 3BFs as opposed to a simple
phenomenological adjustment.  The critical contribution of three-nucleon
interactions has been established in ordinary nuclei in scattering and binding
properties of few-nucleon systems \cite{WGH98,SS02,PPW01} and also in heavier
nuclei and nuclear matter, particularly in connection with saturation
properties \cite{PPW01,APR98,BB12}.

A similar situation is expected in hypernuclei. The possible repulsive effect
from the $\Lambda$NN 3BFs has been speculated
to suppress the appearance of $\Lambda$ hyperons in high-density neutron star
matter by preventing the EoS from becoming soft. Therefore, it is necessary
to quantitatively investigate the effect of the $\Lambda$NN
3BFs in the hypertriton, for which a rigorous treatment is
possible in a Faddeev formulation.

At present, however, the \h3t separation energy 
($B_\Lambda$)
to the deuteron has not been well established experimentally. The previous standard
value of 
$B_\Lambda=130 \pm 50 $ keV
has been updated by the Mainz group
\cite{MA22} by including
recent reports from the STAR and ALICE Collaborations \cite{STAR,ALICE} to
$148\pm 40$ keV. The reproduction of this number, or at least a bound state,
is a test of the 
hyperon-nucleon (YN)
potentials. It is important to know whether the $\Lambda$NN
3BFs provide a non-negligible attractive or repulsive contribution
to tune the description of two-body $\Lambda$N interactions.

While awaiting a more precise value from
the experiments in progress, it is important to quantify
the role of 3BFs for the hypertriton. For this purpose, we perform
Faddeev calculations for the hypertriton, including the $2\pi$-exchange
$\Lambda$NN 3BF in ChEFT, whose expression is given in the
next-to-next-to-leading order (NNLO). In that order, there are
other $1\pi$-exchange and contact terms. However, it is worth considering
the $2\pi$-exchange 3BF for the first attempt to estimate the
contribution of 3BFs, because it has a longer range than other 3BFs and
also is less ambiguous in the sense that the coupling constants are in principle
determined in fitting the NNLO two-body YN interactions. 

The Faddeev equation of the $\Lambda$NN bound state, including $\Lambda$NN
3BFs, is briefly explained in Sec. II-A. The $2\pi$-exchange $\Lambda$NN 3BF
that is taken into account is described in Sec. II-B.
The results calculated using various types of semilocal momentum-space
regularized chiral NN interactions \cite{RKE18} are
presented in Sec. III. A summary follows in Sec. IV.

\section{Faddeev equation of the $\Lambda$NN bound state including 3BFs}
\subsection{Derivation of the Faddeev equation in the presence of $\Lambda$NN force}
The Faddeev equation for a bound $\Lambda$NN system was formulated
and numerical calculations were presented in Ref. \cite{MG93} and also
in Ref. \cite{MK95}. The extension of the Faddeev equation in the presence
of the $\Lambda$NN 3BFs proceeds in the same way as in the case of the
triton with three-nucleon forces \cite{SGS91}.
The derivation of the equation is outlined in the following.

The Schr{\"o}dinger equation for the hypertriton wave function $\Psi$ is
projected into the integral form for a bound state:
\begin{equation}
  \Psi = \frac{1}{E-H_0}(V_{12}+V_{23}+V_{31}+W)\Psi,
\end{equation}
where $H_0$ is a kinetic part of the Hamiltonian of the $\Lambda$NN
system, $V_{ij}$ is a two-body interaction
between the $i$-th and $j$-th baryons, and $W$ represents the 3BF. Hereafter,
the Green function $\frac{1}{E-H_0}$ is denoted by $G_0$ and $i=1$ is assigned to
the $\Lambda$ hyperon. The total wave function $\Psi$ is decomposed into
Faddeev elements, $\Psi =\psi_{12}+\psi_{23}+\psi_{31}$, as
\begin{align}
\psi_{12}=& G_0 V_{12}\Psi,\\
\psi_{23}=& G_0( V_{23}+W)\Psi,\\
\psi_{31}=& G_0 V_{31}\Psi.
\end{align}
Because the $\Lambda$N pair couples to the $\Sigma$N pair, each Faddeev
element has two components with respect to $\Lambda$ and $\Sigma$,
although it is not explicitly written in the above equations.
Introducing two-body $t$-matrices $t_{ij}$ which are defined by the
Lippmann-Schwinger equation
\begin{equation}
t_{ij}=v_{ij}+v_{ij}G_0 t_{ij},
\end{equation} 
these components are verified to satisfy the following three sets of equations.
\begin{align}
 \psi_{12}=& G_0 t_{12}(\psi_{23}+\psi_{31}), \\
 \psi_{23}=& G_0 t_{23}(\psi_{31}+\psi_{12}) \nonumber \\
 &+(G_0+G_0t_{23}G_0)W(\psi_{12}+\psi_{23}+\psi_{31}),\\
 \psi_{31}=& G_0 t_{31}(\psi_{12}+\psi_{23}),
\end{align}
Because the two-nucleon state should be antisymmetric under the permutation
operator $P_{23}$, the relation $\psi_{31}=-P_{23}\psi_{23}$ holds. The 3BF $W$
is included in the second equation, Eq. (7), in which the $\Lambda$ hyperon
is a spectator. Although $W$ may be put in another channel, the present treatment
is more convenient because the expression of the partial-wave expansion of the
2$\pi$-exchange $\Lambda$NN 3BF is simple in this channel \cite{KKM22}.

\subsection{$2\pi$-exchange $\Lambda$NN 3BF}
The $\Lambda$NN 3BFs appear at first in the next-to-next-to-leading
order (NNLO). The $2\pi$-exchange $\Lambda$NN 3BF in momentum space
is given by the diagram shown in Fig. 1 and has a following form \cite{Pet16}.
\begin{align}
 V_{\rm TPE}^{\Lambda {\rm NN}} = \frac{g_A^2}{3f_0^4} (\bftau_2\cdot\bftau_3)
\frac{(\bfsigma_3 \cdot \bq_{3d})(\bfsigma_2 \cdot \bq_{2d})}
{(\bq_{3d}^2+m_\pi^2)(\bq_{2d}^2+m_\pi^2)} \notag \\
 \times \{-(3b_0+b_D)m_\pi^2 +(2b_2+3b_4)\bq_{3d}\cdot\bq_{2d}\},
\label{eq:lnn}
\end{align}
where $\bq_{2d}$ ($\bq_{3d}$) is the difference between the final and initial
momenta at the nucleon line 2 (line 3) of Fig. 1: $\bq_{2d}=\bk_2'-\bk_2$
and $\bq_{3d}=\bk_3'-\bk_3$, and $\bk_i$ are free momentum of the $i$-th particle. 
$g_A$ is the axial coupling constant, $f_0$ is
the pion decay constant, $m_\pi$ is the pion mass, and $\bfsigma_i$
and $\bftau_i$ stand for the spin and isospin operators of the nucleon $i$
(with $i=2,3$), respectively. The coupling constants $b_0$, $b_D$, $b_2$,
and $b_4$ are those in the underlying $\Lambda\Lambda\pi\pi$ Lagrangian.
These coupling constants are to be determined in parametrizing
$\Lambda $N interactions in the next-to-next-to-leading order.
In this paper, we use the values of $3b_0+b_D=0$ and $2b_2+3b_4=-3.0$ GeV$^{-1}$,
following the estimation by Petshauer \textit{et al.} \cite{Pet16} using a decuplet
saturation model. It is noted that the Bonn-J{\"u}lich group has developed
the NNLO parametrization of YN interactions \cite{HMNL23}, in which the
pertinent constants are $3b_0+b_D=-1.485$ and $2b_2+3b_4=-3.01$ GeV$^{-1}$.
Our tentative value of $2b_2+3b_4=-3.0$ GeV$^{-1}$ is reasonable. 
As for $3b_0+b_D$, the possible contribution of this term is commented at the end of Sec. III.

Indeed, here we are focusing solely on the Petshauer coupling constants, but it's worth noting that the subleading meson-baryon LECs that contribute to the NNY 3BF have been extensively discussed and presented in various publications \cite{Mai2009,Liu2007,Ren2012}.

\begin{figure}[t]
\centering
 \includegraphics[width=0.15\textwidth]{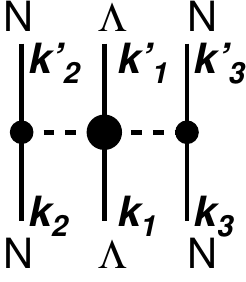}
\caption{Diagram of $2\pi$-exchange $\Lambda$NN 3BF in momentum space.
The small filled circle denotes the ${\rm NN}\pi$ vertex with the coupling
constant $g_A/f_0^2$, and the large filled circle denotes the ${\rm NN}\pi\pi$ vertex
specified by the coupling constants $3b_0+b_D$ and $2b_2+3b_4$ in Eq. (\ref{eq:lnn}).}
\label{fig:lnn}
\end{figure}

In the use of $ V_{\rm TPE}^{\Lambda {\rm NN}}$, the Gaussian regularization function is
introduced with the cutoff scale of $\Lambda_c=550$ MeV, the explicit form is
found in Eq. (4) of Ref. \cite{KKM22}. This prescription is different from the
semilocal regularization for NN potential \cite{RKE18}.
At present, there is no consistent treatment of the cutoff for
the two-body and three-body forces.

In the actual Faddeev calculations, a partial-wave decomposition of
the $\Lambda$NN 3BF is needed. We use the expression developed
in Ref. \cite{KKM22}. In that reference, a $\Lambda$-deuteron folding potential
was evaluated before carrying out actual Faddeev calculations.
As for the NN interaction to prepare the deuteron wave function
in Ref. \cite{KKM22}, the early version of the chiral N$^3$LO interaction \cite{Epe05}
was used. The calculated $\Lambda$-deuteron folding potential was shown to be
weakly attractive and commented that the deuteron wave function is different from
those of other modern NN interactions such as the AV18 \cite{AV18} and
CD-Bonn \cite{CDB}. In the present paper, we employ the recent semilocal
momentum-space regularized chiral NN potentials \cite{RKE18}. For these
chiral NN interactions, the deuteron wave function appears similar to
those of the modern NN interactions mentioned above.
With the new deuteron wave function, the $\Lambda$-deuteron folding potential
turns out to be repulsive, which is illustrated in Fig. 2. Then the  3BF effect
in the hypertriton is expected to be repulsive. The results of
the Faddeev calculations presented below prove this expectation.

\begin{figure}[t]
\centering
 \includegraphics[width=0.4\textwidth]{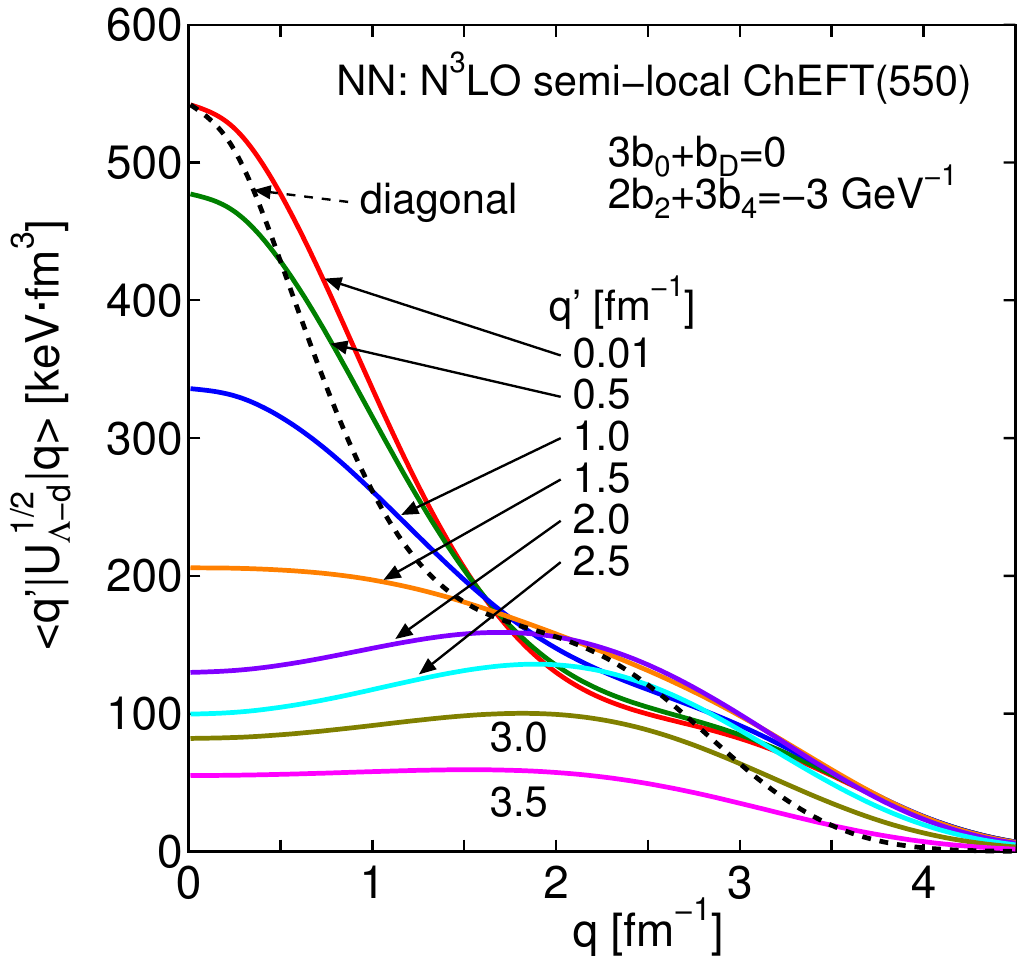}
 \caption{$\Lambda$-deuteron folding potential with $\ell_\Lambda=0$
from $2\pi$-exchange $\Lambda $NN  3BF. The deuteron is
described by the  semilocal N$^3$LO
ChEFT interaction \cite{RKE18} with the cutoff scale of 550 MeV.
The contribution of the deuteron $s$-state pair is shown.}
\label{fig:ldfp}
\end{figure}

\section{Results of Faddeev calculation}
We use four different types of NN interactions constructed in chiral effective
field theory to demonstrate the NN interaction dependence of the \h3t energy;
NLO, N$^3$LO, N$^4$LO, and N$^4$LO+ interactions \cite{RKE18} in which
a semi-local regularization is prescribed. For each type of the NN interaction, four
different values of the cutoff scale, 400, 450, 500, and 550 MeV, are considered.
Altogether, 16 different chiral NN interactions that reproduce
scattering data and the deuteron binding energy equally well are employed. 
Its application to light nuclei and medium-heavy nuclei has also been carried out.\cite{PhysRevC.103.054001,PhysRevC.106.064002}
In numerical evaluations of the Faddeev equation, partial waves up to $j_{max}=4$ are
taken into account in each two-body channel.

As for the $S=-1$ YN interactions, the chiral NLO13 \cite{NLO13} and
NLO19 \cite{NLO19} parameter sets are used, their partial waves up to $j_{max}=2$ are
taken into account in each two-body YN channel.
For comparison, the results of the calculation using the Nigmegen NSC89
potential \cite{NSC89} are presented, although the mixed use of the NN potential
with the cutoff regularization and the YN potential without it may not be consistent.

\begin{widetext}
\begin{center}
\begin{figure}[b]
 \includegraphics[width=0.9\textwidth]{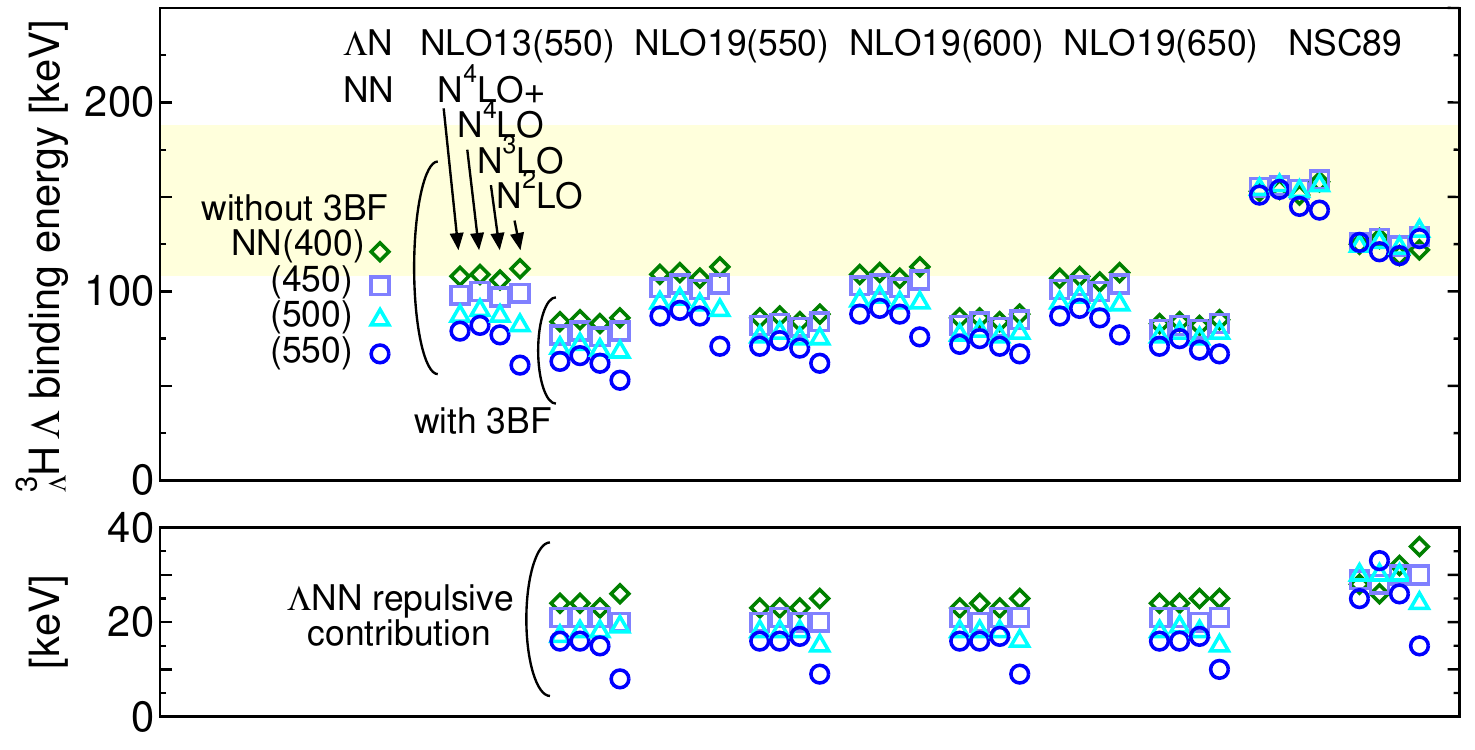}
 \caption{Results of Faddeev calculations with and without $2\pi$-exchange
$\Lambda$NN  3BF using $S=-1$ YN interactions of NLO13, NLO19, and NSC89.
Various NN potentials parametrized in ChEFT are employed. The shaded band
indicates the average in experimental data of $148\pm 40$ keV by the
Mainz group \cite{MA22}. Repulsive contributions of the 3BF are separately
shown in the lower panel.}
\label{fig:fadcal}
\end{figure}
\end{center}
\end{widetext}

Fig. 3 presents the calculated results. Plain numbers are tabulated in Appendix A.
In the upper panel  of Fig. 3, results without and with 3BFs are depicted.
In the absence of the 3BF, the separation energies obtained are around or below 100 keV having
the spreading of the order of 50 keV depending on the cutoff scale of the
NN interaction. The cutoff dependence is somewhat large for N$^2$LO.
Because of the \h3t total isospin $T=0$, isospin singlet NN channels
contribute dominantly, although isospin triplet states enter through the
$\Sigma$NN state. The larger cutoff-momentum tends
to yield slightly smaller separation-energy. On the other hand, the result depends
less on the choice of the chiral YN interaction. The reference values with the
NSC89 YN interaction are about 150 keV and depend
weakly on the NN interactions.

Including the NNLO $2\pi$-exchange $\Lambda$NN  3BF
with $3b_0+b_D=0$ and $2b_2+3b_4=-3.0$ GeV$^{-1}$, the separation
energy decreases. The 3BF effect is repulsive in the order of 20 keV for the
ChEFT $\Lambda$N interactions and 30 keV for the NSC89 $\Lambda$N interaction.
Regarding the small experimental value, this size of the 3BF effect is not negligible.

Before including 3BFs, the calculated separation energy depends on the NN potential
employed, while it is little dependent on the YN interactions used. The spread is
20 - 30 keV. Similar uncertainties in the hypertriton system are reported in the literature
\cite{HGFY21, GHF22}. The NN potential with the larger cutoff-momentum
tends to yield smaller separation energy. This trend agrees with the results
in Ref. \cite{HGFY21,GHF22}.
The small separation-energy means that the wave function of the system
is spreading wider compared with the one bearing larger separation-energy.
Then, the repulsive effect of the 3BF is expected to be smaller for the former than for the
latter. Consequently, the spread of the separation energy depending on the
NN cutoff scale is slightly reduced after including 3BFs.
The 3BF contribution is shown separately in the lower part of Fig. 3.
Although some irregularity is seen for the reference results using the
NSC89 YN interaction, these results demonstrate that the 3BF effect is
larger for the YN interaction that yields a larger \h3t separation energy.
 
In the present Faddeev calculations, the coefficient $(3b_0+b_D)$ in Eq. (9) is set
to be 0. It is instructive to comment on the possible effect of this term.
It is understood from Eq. (12) of Ref. \cite{MK18}, which shows the contribution
of the $2\pi$-exchange 3BF to the $\Lambda$ single-particle potential
in symmetric nuclear matter, that if $(3b_0+b_D)$ and $(2b_2+3b_4)$ have a
same magnitude with the same sign, the term of $(3b_0+b_D)$
provides about one-third of the contribution of that
of $(2b_2+3b_4)$. This character probably holds also in the hypertriton.
Including the term with non-zero $(3b_0+b_D)$ in the Faddeev calculation
is straightforward when the coupling constant becomes under control.

\section{Summary}
We have performed Faddeev calculations of the hypertriton separation energy
incorporating the 2$\pi$-exchange $\Lambda$NN 3BF and estimated
the effect of the 3BF. The contribution is found repulsive in the order of $20$ keV,
although we have to bear in mind that the amount depends on the sign
and strength of the coupling constants. The obtained repulsive magnitude
is small, as is expected from the loosely bound nature of \h3t. However, it cannot
be neglected regarding the small experimental separation energy.
In other words, the 2$\pi$-exchange $\Lambda$NN 3BF has some degree of impact. By carefully adjusting the $S=-1$ YN interaction while considering this influence, we could provide a more accurate description of the hypertriton separation energy once more precise experimental data becomes available.

Based on the present experimental average,  it is seen that the available chiral YN two-body interactions NLO13 and NLO19 give a smaller amount of 
separation energies when the $2\pi$-exchange 3BF is taken into account with the provisional coupling constants. Naturally, the remaining NNLO 3BFs,
$1\pi$-exchange and contact terms, need to be considered. Those studies
are in progress. Calculations in other light hypernuclei are also required to determine
the relevant coupling constants. The settlement of the coupling constants of
$\Lambda $NN 3BFs is {important} to gauge the contribution
of the 3BFs to the $\Lambda$ single-particle energy in neutron star matter,
which relates to the hyperon puzzle.

\medskip
{\it Acknowledgements.}
We are grateful to J. Haidenbauer for providing us the program code and
parameters of chiral NLO YN interactions. This work is supported by
JSPS KAKENHI Grants No. JP19K03849 and No. JP22K03597.

\appendix
\section{Numerical results}
Results of the calculations of the \h3t separation energy are depicted in Fig. 3
in Sec. III.  Plain numbers are tabulated in this Appendix.
Table I shows the deuteron energy generated by each semilocal chiral NN potential\cite{RKE18}.
The $_\Lambda^3$H separation energy refers to this value for each NN potential.
Tables II and III represent the calculated hypertriton separation energy $B_\Lambda$ with and
without $2\pi$-exchange $\Lambda$NN 3BFs, respectively.
In Table III of Ref. \cite{NLO19}, the separation energies of $_\Lambda^3$H for 
NLO19(650) with NN interactions N$^4$LO(400), N$^4$LO(450), N$^4$LO(500),
and N$^4$LO(550) are presented. The numbers in Table II agree with those within
10 keV. 

In Tables II and III, some calculated values for other realistic NN potentials 
(CDBonn\cite{CDB}, Nijmegen\cite{Nijm93}) are added for reference.

\begin{table}[b]
\caption{\label{tab1} Deuteron bound energies of the semilocal chiral NN
potentials \cite{RKE18} employed. Entries are in MeV.
The numbers in parentheses indicate the cutoff scale $\Lambda_c$ in MeV.}
{
\begin{ruledtabular}
\begin{tabular}{cc} 
Chiral Order ($\Lambda_c$) & Deuteron Energy\\ \hline
N$^4$LO+ (550) & $2.2230$ \\
N$^4$LO+ (500) & $2.2230$ \\
N$^4$LO+ (450) & $2.2230$ \\
N$^4$LO+ (400) & $2.2230$ \\
\hline
N$^4$LO  (550) & $2.2230$ \\
N$^4$LO  (500) & $2.2230$ \\
N$^4$LO  (450) & $2.2230$ \\
N$^4$LO  (400) & $2.2230$ \\
\hline
N$^3$LO  (550) & $2.2230$ \\
N$^3$LO  (500) & $2.2230$ \\
N$^3$LO  (450) & $2.2230$ \\
N$^3$LO  (400) & $2.2230$ \\
\hline
N$^2$LO  (550) & $2.2416$ \\
N$^2$LO  (500) & $2.2188$ \\
N$^2$LO  (450) & $2.1997$ \\
N$^2$LO  (400) & $2.1792$ 
\end{tabular}
\end{ruledtabular}
}

\end{table}

\begin{table}
\caption{\label{tab2} Hypertriton separation energies without 3BFs.
Entries are in keV. The numbers in parentheses for NN and YN chiral
potentials indicate the cutoff scale $\Lambda_c$ in MeV. }
{
\begin{ruledtabular}
\begin{tabular}{c|ccccc} 
    & \multicolumn{5}{c}{YN potential ($\Lambda_c$)  } \\ 
NN potential  & NLO13& NLO19& NLO19& NLO19&NSC89\\
 ~~~~~~~($\Lambda_c$)  & (550)& (550)& (600)& (650)&      \\   \hline
CDBonn      \cite{CDB}  & $\phantom{0}82$ &  $\phantom{0}90$ &  $\phantom{0}90$
                 &  $\phantom{0}88$ & $153$ \\ 
Nijmegen 93 \cite{Nijm93}  & $\phantom{0}58$ &  $\phantom{0}69$ &  $\phantom{0}70$
                &  $\phantom{0}70$ & $141$ \\
Nijmegen I  \cite{Nijm93}  & $\phantom{0}63$ &  $\phantom{0}75$ &  $\phantom{0}75$
               &  $\phantom{0}74$ & $145$ \\ \hline
N$^4$LO+ (550) & $ \phantom{0}79$ & $ \phantom{0}87$ & $ \phantom{0}88$
                       & $ \phantom{0}87$ & $151$ \\
N$^4$LO+ (500) & $ \phantom{0}87$ & $ \phantom{0}94$ & $ \phantom{0}95$
                       & $ \phantom{0}94$ & $154$ \\
N$^4$LO+ (450) & $ \phantom{0}98$ & $102$ & $103$ & $101$ & $155$ \\
N$^4$LO+ (400) & $108$ & $109$ & $109$ & $107$ & $153$ \\
\hline
N$^4$LO  (550) & $ \phantom{0}82$ & $ \phantom{0}90$ & $ \phantom{0}91$
                      & $ \phantom{0}91$ & $154$ \\
N$^4$LO  (500) & $ \phantom{0}90$ & $ \phantom{0}96$ & $ \phantom{0}97$
                      & $ \phantom{0}97$ & $156$ \\
N$^4$LO  (450) & $100$ & $104$ & $104$ & $103$ & $156$ \\
N$^4$LO  (400) & $109$ & $110$ & $110$ & $108$ & $154$ \\
\hline
N$^3$LO  (550) & $ \phantom{0}77$ & $ \phantom{0}87$ & $ \phantom{0}88$
                      & $ \phantom{0}86$ & $145$ \\
N$^3$LO  (500) & $ \phantom{0}87$ & $ \phantom{0}93$ & $ \phantom{0}94$
                      & $ \phantom{0}93$ & $153$ \\
N$^3$LO  (450) & $ \phantom{0}97$ & $101$ & $102$ & $100$ & $154$ \\
N$^3$LO  (400) & $106$ & $107$ & $107$ & $105$ & $151$ \\
\hline
N$^2$LO  (550) & $ \phantom{0}61$ & $ \phantom{0}71$ & $ \phantom{0}76$
                      & $ \phantom{0}77$ & $143$ \\
N$^2$LO  (500) & $ \phantom{0}82$ & $ \phantom{0}90$ & $ \phantom{0}94$
                      & $ \phantom{0}93$ & $156$ \\
N$^2$LO  (450) & $ \phantom{0}99$ & $104$ & $106$ & $104$ & $159$ \\
N$^2$LO  (400) & $112$ & $113$ & $113$ & $110$ & $158$ \\  
\end{tabular}
\end{ruledtabular}
}

\end{table}

\begin{table}
\caption{\label{tab3} Hypertriton separation energies including 2$\pi$-exchange
$\Lambda $NN  3BF with $3b_0+b_D=0$ and $2b_2+3b_4={-3.0}$ GeV$^{-1}$.
Entries are in keV. The numbers in parentheses for NN and YN chiral
potentials indicate the cutoff scale $\Lambda_c$ in MeV.}
\begin{ruledtabular}
\begin{tabular}{c|ccccc}
    & \multicolumn{5}{c}{YN potential ($\Lambda_c$)  } \\ 
 NN potential               & NLO13& NLO19& NLO19& NLO19& NSC89\\
 ~~~~~~~~($\Lambda_c$)  & (550)& (550)& (600)& (650)&  \\\hline

 CDBonn     \cite{CDB}      & $ 66$  & $ 73$  & $ 72$  & $ 71$  & $126$\\  
Nijmegen 93 \cite{Nijm93}   & $ 48$  & $ 58$  & $ 58$  & $ 59$  & $121$\\ 
Nijmegen I  \cite{Nijm93}   & $ 51$  & $ 62$  & $ 63$  & $ 62$  & $123$\\
 \hline
N$^4$LO+ (550)    & $ 63$  & $ 71$  & $ 72$  & $ 71$  & $120$\\
N$^4$LO+ (500)    & $ 70$  & $ 76$  & $ 77$  & $ 76$  & $124$\\
N$^4$LO+ (450)    & $ 77$  & $ 82$  & $ 82$  & $ 80$  & $126$\\
N$^4$LO+ (400)    & $ 84$  & $ 86$  & $ 86$  & $ 83$  & $125$\\
 \hline
N$^4$LO  (550)    & $ 66$  & $ 74$  & $ 75$  & $ 75$  & $121$\\
N$^4$LO  (500)    & $ 72$  & $ 78$  & $ 79$  & $ 78$  & $126$\\
N$^4$LO  (450)    & $ 79$  & $ 83$  & $ 84$  & $ 82$  & $128$\\
N$^4$LO  (400)    & $ 85$  & $ 87$  & $ 86$  & $ 84$  & $128$\\
 \hline
N$^3$LO  (550)    & $ 62$  & $ 70$  & $ 71$  & $ 69$  & $119$\\
N$^3$LO  (500)    & $ 69$  & $ 75$  & $ 76$  & $ 75$  & $123$\\
N$^3$LO  (450)    & $ 76$  & $ 81$  & $ 81$  & $ 80$  & $124$\\
N$^3$LO  (400)    & $ 83$  & $ 84$  & $ 84$  & $ 82$  & $119$\\
 \hline
N$^2$LO  (550)    & $ 53$  & $ 62$  & $ 67$  & $ 67$  & $ 128 $\\
N$^2$LO  (500)    & $ 68$  & $ 75$  & $ 78$  & $ 78$  & $ 132 $\\
N$^2$LO  (450)    & $ 79$  & $ 84$  & $ 85$  & $ 83$  & $ 129 $\\
N$^2$LO  (400)    & $ 86$  & $ 88$  & $ 88$  & $ 85$  & $ 122 $\\
\end{tabular}\end{ruledtabular}

\end{table}

\end{document}